\documentclass[graybox]{svmult}

\usepackage{hyperref}
\usepackage{cite}

\usepackage{mathptmx}       
\usepackage{helvet}         
\usepackage{courier}        
\usepackage{type1cm}        
\usepackage{makeidx}         
\usepackage{graphicx}        
\usepackage{multicol}        
\usepackage[bottom]{footmisc}

\usepackage{amsmath}
\usepackage{url}

\makeindex             

\begin{document}

\title*{Modelling the unfolding pathway of biomolecules: theoretical approach and experimental prospect}
\titlerunning{Modelling the unfolding pathway of biomolecules} 

\author{Carlos A.~Plata and Antonio Prados}
\institute{Carlos A.~Plata \at F\'{\i}sica Te\'{o}rica, Universidad de Sevilla,
Apartado de Correos 1065, E-41080 Seville, Spain, EU, \email{cplata1@us.es}
\and Antonio Prados \at F\'{\i}sica Te\'{o}rica, Universidad de Sevilla,
Apartado de Correos 1065, E-41080 Seville, Spain, EU, \email{prados@us.es}}
%
%
\maketitle

\abstract*{We analyse the unfolding pathway of biomolecules comprising
  several independent modules in pulling experiments. In a recently
  proposed model, a critical velocity $v_{c}$ has been predicted, such
  that for pulling speeds $v>v_{c}$ it is the module at the pulled end
  that opens first, whereas for $v<v_{c}$ it is the weakest. Here, we
  introduce a variant of the model that is closer to the experimental
  setup, and discuss the robustness of the emergence of the critical
  velocity and of its dependence on the model parameters. We also
  propose a possible experiment to test the theoretical predictions of
  the model, which seems feasible with state-of-art molecular
  engineering techniques. }

\abstract{We analyse the unfolding pathway of biomolecules comprising
  several independent modules in pulling experiments. In a recently
  proposed model, a critical velocity $v_{c}$ has been predicted, such
  that for pulling speeds $v>v_{c}$ it is the module at the pulled end
  that opens first, whereas for $v<v_{c}$ it is the weakest. Here, we
  introduce a variant of the model that is closer to the experimental
  setup, and discuss the robustness of the emergence of the critical
  velocity and of its dependence on the model parameters. We also
  propose a possible experiment to test the theoretical predictions of
  the model, which seems feasible with state-of-art molecular
  engineering techniques. }


\section{Introduction}
\label{sec:intro}
The development of the so-called single-molecule experiments in the
last decades has made it possible to carry out research at the
molecular level. Biophysics is, undoubtedly, one of the fields where
these techniques have had a bigger impact, triggering a whole new area
of investigation on the elastic properties of biomolecules. Recent
accounts of the current development of this enticing field can be
found in Refs.~\cite{R06,KyL10,MyD12,HyD12}.

\begin{figure}
  \centering
 \includegraphics[width=0.85 \textwidth]{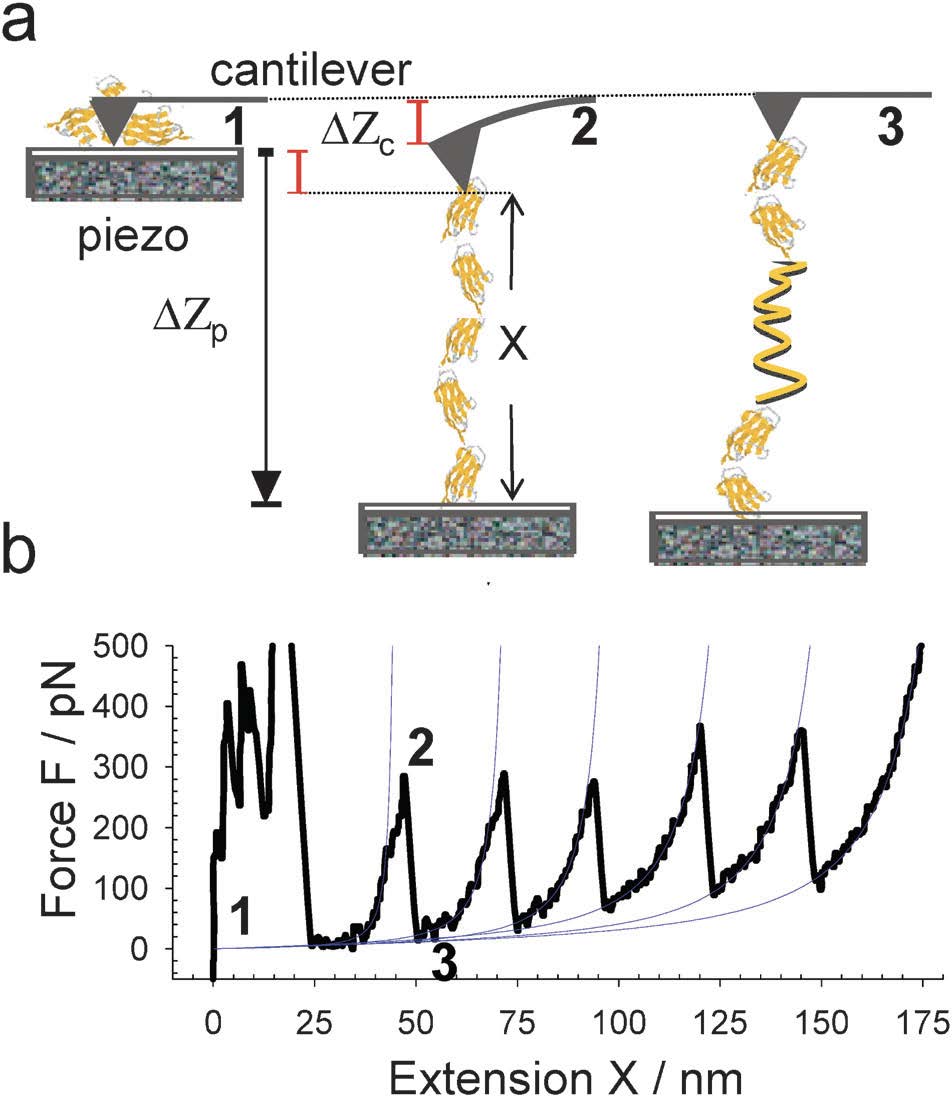}
 \caption{\textbf{a}: Sketch of the experimental setup in an AFM
   experiment with a modular protein. The position of the platform is
   shifted $\Delta Z_{p}$ from 1 to 2, producing an elongation of $X$
   over the molecule and bending the cantilever a magnitude
   $\Delta Z_{c}$. From 2 to 3 the force is almost relaxed because of
   the unravelling of one of the modules. \textbf{b}: Force-extension
   curve in a typical lenght-control AFM experiment with a
   polyprotein. Each rip in the force accounts for the unfolding of a
   module. Taken from \cite{MyD12}.}
  \label{fig:sketch-experiment}
\end{figure}

Atomic force microscopy (AFM) stands out because of its extensive
use. In particular, the role played by AFM in the study of modular
proteins is crucial
\cite{COFMBCyF99,FMyF00,HDyT06}. Figure~\ref{fig:sketch-experiment}
shows a sketch of the experimental setup in a pulling experiment of a
molecule comprising several modules. The biomolecule is stretched
between the platform and the tip of the cantilever. The spring
constant of the cantilever is $k_{c}$, which is usually in the range
of $100$ pN/nm. Here, we consider the simplest situation, in which the
total length of the system $\Delta Z_{p} = \Delta Z_{c}+X$ is
controlled. The stretching of the molecule makes the cantilever bend
by $\Delta Z_{c}$, and then the force can be recorded as
$F=k_{c}\Delta Z_{c}$.

The outcome of the above described experiment is a force-extension
curve, similar to panel b) in Fig.~\ref{fig:sketch-experiment}. This
force-extension curve provides a fingerprint of the elastomechanical
properties of the molecule under study. When molecules composed of
several structural units, such as modular polyproteins, are pulled, a
sawtooth pattern comes about in the force-extension curve
\cite{COFMBCyF99,FMyF00,HDyT06}. At certain values of the length,
there are almost vertical ``force rips'': each force rip marks the
unfolding of one module. Interestingly, these force rips already
appear when the molecule is quasi-statically pulled, a limit that can
be explained by means of an equilibrium statistical mechanics
description \cite{PCyB13}. When the molecule is pulled at a finite
rate, the appearance of these force rips can still be explained by the
system partially sweeping the metastable region of the equilibrium
branches \cite{BCyP15}.

The unfolding pathway is, roughly, the order and the way in which the
structural units of the system unravel. It has been recently found out
that the unfolding pathway depends on the pulling velocity and the
pulling direction \cite{HDyT06,LyK09,GMTCyC14,KHLyK13}. Particularly,
in \cite{GMTCyC14}, different unfolding pathways are observed in SMD
simulations on the Maltose Binding Protein. The authors reported that
for low pulling speeds the first unit to unfold is the least stable,
whereas for high pulling speed it is the closest to the pulled end,
regardless of their relative stability.

Very recently, a toy-like model has been proposed to qualitatively
understand the above experimental framework
\cite{PCCyP15}. Specifically, each module is represented by a
nonlinear spring, characterised by a bistable free energy that depends
on the module extension. Therein, the two basins represent its folded
and unfolded states. The spatial structure of the system is retained
in its simplest way: each module extends from the end point of the
previous one to its own endpoint (which coincides with the start point
of the next). Moreover, each module endpoint obeys an overdamped
Langevin equation with  forces stemming from the bistable free
energies and white noise forces with amplitudes verifying the
fluctuation-dissipation theorem.

In the above model, the unfolding pathway was found to depend on the
pulling velocity. In the simplest non-trivial case, there is only one
module that is different from the rest, which is also the furthest
from the pulled end.  In this situation, only one critical velocity
$v_{c}$ shows up: for pulling velocities $v_{p}>v_{c}$, it is the
weakest module that opens first but for $v_{p}>v_{c}$ it is the module
at the pulled end. In addition, analytical results were derived for
this critical velocity by introducing some approximations: mainly two,
(i) perfect length control and (ii) the deterministic approximation,
that is, our neglecting of the stochastic forces. This was done by
means of a perturbative solution of the deterministic equations in
both the pulling velocity and an asymmetry parameter, which measures
how different the potentials of the modules are.

The main aim of this work is twofold. First, we would like to refine
the above theoretical framework, making it closer to the experimental
setup in AFM. In particular, we would like to look into the effect of
a more realistic modelling of the length-control device. Instead of
considering perfect length-control, we consider a device with a finite
value of the stiffness, both at the end of the one-dimensional chain
(as originally depicted in Ref.~\cite{PCCyP15}) and at the start point
thereof, which is where it is usually situated in the AFM experiments,
see Fig.~\ref{fig:sketch-experiment}. Second, we would like to discuss
how our theory could be checked in a real experiment with modular
proteins.

This chapter is structured as follows. First, we introduce the
original model and discuss its most relevant results in section
\ref{sec:orig}. In section \ref{sec:stiff}, we study the role played
by the location of the restoring spring and the finite value of its
spring constant. We provide some details about the free energy
modelling employed for each of the modules in section
\ref{sec:potentials}. Section \ref{sec:experiments} is devoted to
discuss a possible AFM experiment in order to test our
theory. Finally, we wrap up the main conclusions which emerge from our
work.

\section{Revision of the model and previous results}
\label{sec:orig}

\begin{figure}
\centering
  \includegraphics[width=0.85 \textwidth]{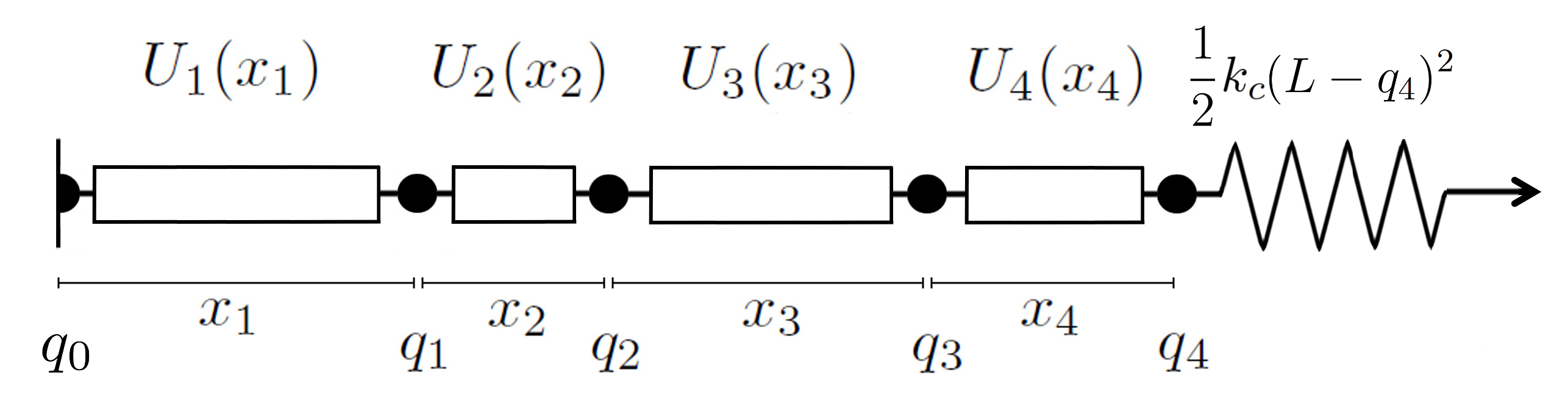}
  \caption{\label{fig:1} Sketch of the model for a molecule comprising
    four units. Therein, the units are denoted by rectangles and have
    potentials $U_i(x_i)$, with $x_i$ being the extension of the
    $i$-th unit's. The unit endpoints $q_{i}$ are represented by the
    beads, and the extensions are thus $x_{i}=q_{i}-q_{i-1}$ (by
    definition, $q_{0}=0$). The spring stands for the
    length-controlling device attached to the pulled end $q_4$, whose
    contribution to the system free energy is assumed to be harmonic
    with stiffness $k_{c}$.
  }
\end{figure}

Here, we briefly review the model that was originally put
  forward in Ref.~\cite{PCCyP15}. We consider a polyprotein
comprising $N$ modules. When the molecule is stretched, the simplest
description is to portray it as a one-dimensional chain. We define the
coordinates $q_i,$ $(i=0,\ldots,N)$ in such a way that the $i$-th unit
extends from $q_{i-1}$ to $q_i$, the extension of the $i$-th unit is
$x_i=q_i-q_{i-1}$. Moreover, as shown in Fig.~\ref{fig:1}, the pulling
device is assumed to be connected to the right (pulled)  end of the chain.
 	
We assume Langevin dynamics for the $q_i$ coordinates ($q_0=0$),
\begin{equation}
\label{langevin}
  \gamma \dot{q}_{i}=-\frac{\partial}{\partial q_{i}} A(q_{0},\ldots,q_{N})+\zeta_{i} \qquad i>0 ,
\end{equation}
in which the $\zeta_{i}$ are Gaussian white noise forces, such that
\begin{equation}
\langle \zeta_{i}(t)\rangle=0, \qquad 
\langle \zeta_{i}(t) \zeta_{j}(t')\rangle=2 \gamma k_{B} T \delta_{ij}
\delta(t-t'),
\end{equation}
with $k_B$ being the Boltzmann constant, and $\gamma$ and $T$ being
the friction coefficient (assumed to be common for all the units) and
the temperature of the fluid in which the protein is immersed,
respectively.  The global free energy function of the system
is\footnote{In Ref.~\cite{PCCyP15}, this free energy was denoted by
  $G$. Here, we have preferred to employ $A$ because
  the relevant potential in length-controlled situations is the
  Hemlholtz-like free energy, and $G$ is usually the notation reserved
  for the Gibbs-like potential $G=A-FL$, which is the relevant one in
  force-controlled experiments \cite{PCyB13,BCyP15}.}
\begin{equation}
\label{freeG}
  A(q_{0},\ldots,q_{N})=\sum_{i=1}^{N} U_{i}(q_{i}-q_{i-1})+\frac{1}{2}k_c (L-q_N)^2.
\end{equation}
In the previous equation, we have considered an elastic term due to
the finite stiffness $k_c$ of the controlling device, which is located
at the pulling end as shown in Fig.~\ref{fig:1}. Finally, $L$ stands
for the desired length program and $U_i(x_i)$ is the single unit
contribution, which only depends on the extension, to
$A$. Consequently, the force exerted over the biomolecule is
$k_c(L-q_N)$. We consider length-controlled experiments at constant
pulling velocity, that is, $\dot{L}=v_p$.

When $k_c \to \infty$, the control over the length is perfect and $q_N \to L$ in such a way that
$k_c (L-q_N) \to F$, being $F$ a Lagrange multiplier. That is, the
perfect length-controlled situation is the same that a
force-controlled one but with $F$ the force needed to maintain a total
length equal to $L$. Note that there is no contribution to the
free energy coming from the pulling device in this limit, since
$k_{c}(L-q_{N})^{2}/2=F^{2}/(2k_{c})\to 0$.

The approach in Ref.~\cite{PCCyP15} tries to keep things as
simple as possible. Then, the evolution equations for the extensions
are written by assuming (i) perfect length control and
(ii) the deterministic approximation, obtained by neglecting
  the noise terms. Note that the evolution equations in the latter
approximation are sometimes called the macroscopic equations
\cite{vK92}, which are
\begin{subequations}
\label{eq:7}
\begin{align}
\gamma\dot{x}_{1}  &=  -U'_{1}(x_{1})+U'_{2}(x_{2}),\\
\gamma\dot{x}_{i}  &=  -2U'_{i}(x_{i})+U'_{i+1}(x_{i+1})+U'_{i-1}(x_{i-1}),
     \quad 1<i<N,\\
  \gamma\dot{x}_{N}  &=  -2U'_{N}(x_{N})+U'_{N-1}(x_{N-1})+F, \\
  F &= \gamma v_{p}+U'_{N}(x_{N}).
\end{align}
\end{subequations}

So far, nothing has been said about the shape of the single-unit
contributions $U_i$ to the free energy. In order to maintain a general
approach, we only request the functions $U_i(x)-F x$ to become a
double well for some interval of forces. Each well stands for the
folded and the unfolded basins of each module.  Now, we separate these
functions in a main part common to all units and a separation from
this main part weighted by an asymmetry parameter $\xi$,
\begin{equation}
\label{as_sep}
U_i(x)=U(x)+\xi \delta U_i(x), \quad U'_i(x)=U'(x)+\xi \delta f_i(x).
\end{equation}
We have done the same separation in the derivative, by defining $\delta f_i(x) = \delta U'_i(x)$.

It is possible to solve the system \eqref{eq:7} by means of a perturbative expansion in the pulling velocity
$v_p$ and the asymmetry $\xi$. Indeed, if we retain only linear order
terms in $v_p$ and $\xi$, the corrections due to finite pulling rate
and asymmetry are not coupled. This perturbative solution, when the
system starts from an initial condition in which all the units are
folded, is shown to be \cite{PCCyP15}
\begin{equation}
x_{i}=\ell+ \frac{\xi\overline{\delta f}-v_{p}\gamma\dfrac{N^2-1}{6N}}{U''(\ell)}
   + \frac{v_{p}\gamma \dfrac{ i(i-1)}{2N}-\xi \delta f_{i}(\ell)}{U''(\ell)},
\label{eq:23}
\end{equation}
where $\ell=L/N$ stands for the specific length per module and the
over-bar means average over the units. Note that, if all the free
energies are equal ($\xi=0$) and we are not pulling ($v_p=0$), the
total length will be reasonably equally distributed among all the
units. Moreover, it is worth emphasising that this solution is
approximate, it diverges when $U''(\ell) \to 0$. This shows that this
perturbative solution breaks down when the average length per module
$\ell$ reaches the \textit{stability threshold} $\ell_{b}$, such that
$U''(\ell_{b})=0$.

  We are interested in a criterion that allows us to discern which
  unit is the first to unfold and we hope that our perturbative
  solution is good enough in this regard. Since the folded state
  ceases to exist when $x$ reaches $\ell_b$, it is reasonable to
  assume that the first module to unfold is precisely the one for
  which $x_i=\ell_b$ is attained for the shortest time. In
  Eq.~\eqref{eq:23}, we can see that the finite pulling term favours
  the unfolding of modules that are nearer to the pulled end, whereas
  the asymmetry term favours the unfolding of the weaker units (those
  with the lowest values of $\delta f_i$).

We can compute the pulling velocities $v^{i}(j)$ for which each couple of modules $(i,j)$, $j>i$, reach
  simultaneously the stability threshold. They are determined by the
  condition
\begin{equation}
  \label{eq:stab}
  x_{i}(\ell_{c})=x_{j}(\ell_{c}) = \ell_{b},
\end{equation}
which gives both the value of $\ell_{c}$
(or time $t_{c}$) at which the stability threshold is reached and the
relationship between $v_{p}$ and $\xi$. Then, in a specific system with known $\delta f_i$'s we
can predict what are the critical velocities that separate regions
inside which the first unit to unfold is different. In
Ref.~\cite{PCCyP15}, some examples of the use of this theory
are provided, which show a good agreement with
simulations of the Langevin dynamics \eqref{langevin}.

The simplest configuration in which a critical velocity arises is the
following.  Let us consider a chain of $N$ units, all of them with the
same contribution to the free energy, except the first one (the
furthest to the pulled end). Therefore, $\delta f_i(x)=0$, $i\neq 1$,
$\delta f_1(x)\neq 0$. Moreover we will assume that
$\delta f_1(\ell_b)<0$, that is, the first module is weaker than the
rest.  For such a configuration, there appears only one critical
velocity, which is given by \cite{PCCyP15}
\begin{equation}
\label{eq:vc}
\frac{\gamma v_c}{\xi} = - \frac{\delta f_1(\ell_b)}{N-1}.
\end{equation}
For $v_p<v_c$, the first module to unfold is the weakest one, whereas
for $v_p>v_c$ the unfolding starts from the pulled end. In
Ref.~\cite{PCCyP15}, a more general situation is investigated but,
here, we restrict ourselves to this configuration.

\section{Relevance of the stiffness}
\label{sec:stiff}
\begin{figure}
\centering
  \includegraphics[width=0.85 \textwidth]{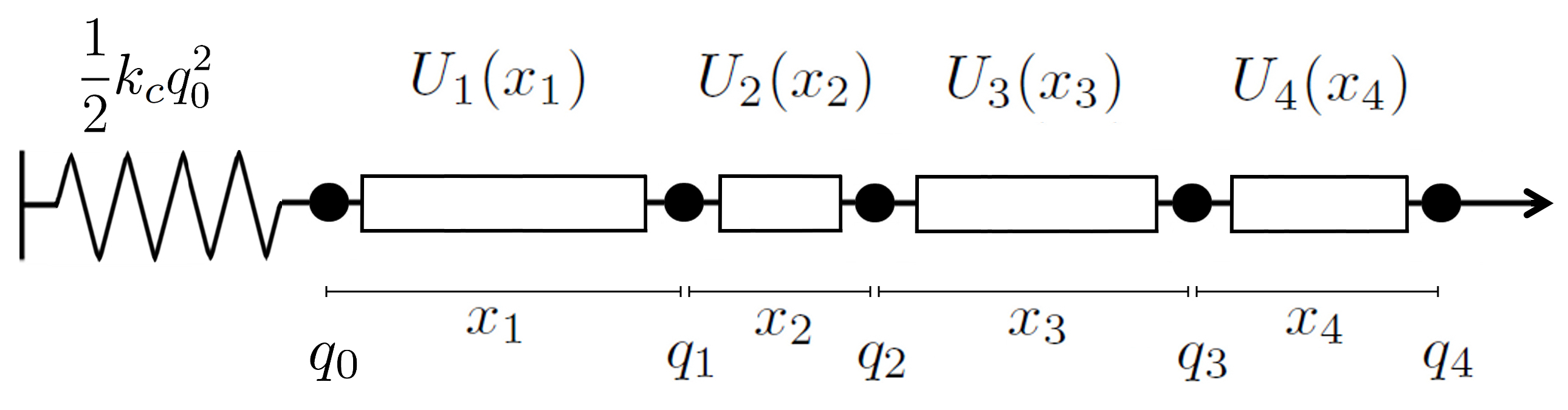}
  \caption{\label{fig:2} Sketch of the model for a protein with four
    units. It is identical toFig.~\ref{fig:1}, except for the position
    of the length-controlling device, which is now located at the
    fixed end.  }
\end{figure}

In a real AFM experiment, the stiffness is finite and, as a result,
the control over the length is not perfect. Furthermore, the position
that is externally controlled is, usually, that of the platform and
the main elastic force stems from the bending of the tip of the
cantilever, as depicted in Fig.~\ref{fig:sketch-experiment}. Thus, it
seems more reasonable to model the pulling of the biomolecule in the
way sketched in Fig.~\ref{fig:2}.

Some authors \cite{HyS03} have used other elastic reactions that
reflect the attachment by means of flexible linkers among the platform
and the pulled end $q_N$, and between consecutive modules. Here, we
will consider a perfect absorption, in order to keep the model as
simple as possible. In the next two subsections, we study the effect
of the finite value of the stiffness $k_c$ and the location of the
spring, respectively, on the unfolding pathway.

\subsection{Finite stiffness}

Here, we still consider the model depicted in
  Fig.~\ref{fig:1}, that is, the spring is situated at the end of the
  chain, but with a finite value of the stiffness
$k_c$. Also, we consider the macroscopic equations (zero
  noise), which are
\begin{subequations}
\begin{align}
\label{eq:7v2}
  \gamma\dot{x}_{1}  &=  -U'_{1}(x_{1})+U'_{2}(x_{2}), \\
  \gamma\dot{x}_{i}  &=  -2U'_{i}(x_{i})+U'_{i+1}(x_{i+1})+U'_{i-1}(x_{i-1}),
     \quad 1<i<N, \\
  \gamma\dot{x}_{N}  &=  -2U'_{N}(x_{N})+U'_{N-1}(x_{N-1})+ k_c \left(L- \sum_{k=1}^{N} x_k \right).
\end{align}
\end{subequations}
This system differs from that in Eq.~\eqref{eq:7}
because in the last equation the Lagrange multiplier
$F$ is substituted by the harmonic force $k_c(L- \sum_{k=1}^{N} x_k)$. As in the previous case,
this system is analytically solvable by means of a perturbative
expansion in $v_p$ and $\xi$. The approximate solution for the
extension $x_i$ is
\begin{eqnarray}
x_{i}=\ell &+& \frac{\xi N k_c \overline{\delta f}(\ell)-v_{p}\gamma k_c \dfrac{[3U''(\ell)+k_c(N-1)]N(N+1)}{6[Nk_c+U''(\ell)]}}{U''(\ell)[Nk_c+U''(\ell)]}  \nonumber
\\
&+& \frac{v_{p}\gamma k_c i(i-1)-2\xi [Nk_c+U''(\ell)] \delta f_{i}(\ell)}{2U''(\ell)[Nk_c+U''(\ell)]}.
\label{eq:24}
\end{eqnarray}
Here $\ell \neq L/N$, it stems from the relation
\begin{equation}
\label{zeroth}
U'(\ell)=k_c(L-N\ell).
\end{equation}
We can see easily how we reobtain \eqref{eq:23} taking the limit
$k_c \to \infty$ in \eqref{eq:24}, as it should be. Although the
solution is slightly different, it still breaks down when $U''(\ell)$
vanishes, that is, when $\ell\to \ell_{b}$. Therefore, to the lowest
order, again we have to seek a solution of \eqref{eq:stab}, with the
extensions given by \eqref{eq:24}, and substitute
$\ell_c \simeq \ell_b$ therein. This leads to the same critical
velocities found for the infinite stiffness limit.

\subsection{Location of the elastic reaction}

As depicted in Fig.~\ref{fig:sketch-experiment}, in an AFM experiment
the distance between the moving platform and the fixed cantilever is
the controlled quantity.  Then, the model sketched in Fig.~\ref{fig:2}
is closer to the experimental setup: the left end corresponds to the
fixed cantilever, with $q_0$ standing for $\Delta Z_{c}$, and the
right end represents the moving platform. Thus, the free energy of
this setup is given by
\begin{equation}
\label{freeG2}
  A(q_{0},\ldots,q_{N})=\sum_{i=1}^{N} U_{i}(q_{i}-q_{i-1})+\frac{1}{2}k_c q_0^2 \, .
\end{equation}

From the free energy \eqref{freeG2}, we derive the Langevin
equations by making use of Eq.~\eqref{langevin}. The macroscopic
equations (zero noise) read
\begin{subequations}\label{eq:7v3}
\begin{align}
  \gamma\dot{x}_{1}  &=  -2 U'_{1}(x_{1})+U'_{2}(x_{2})+ k_c \left(L- \sum_{k=1}^{N} x_k \right), \\
  \gamma\dot{x}_{i}  &=  -2U'_{i}(x_{i})+U'_{i+1}(x_{i+1})+U'_{i-1}(x_{i-1}),
     \quad 1<i<N, \\
  \gamma\dot{x}_{N}  &=  -U'_{N}(x_{N})+U'_{N-1}(x_{N-1})+ v_p.
\end{align}
\end{subequations}
In the infinite stiffness limit, $k_c \to \infty$, the
harmonic contribution tends to a new Lagrange multiplier $F$
such that $\sum x_{i}=L$. Therefrom, it is obtained that
$F=U_1'(x_1)$ and the resulting system is exactly equal to
  that in Eq.~\eqref{eq:7}. This is logical: if the spring
is totally stiff and then the control over the length is perfect, the
two models are identical.\footnote{It is worth emphasising that
  the two variants of the model, with the spring at either the
    fixed or moving end, have the same number of degrees of
  freedom. In Fig.~\ref{fig:1}, $q_0=0$ and our degrees of freedom are
  $q_i$, $i=1, \ldots, N$, whereas in Fig.~\ref{fig:2} we have the
  dynamical constraint $q_N=L$ and the degrees of freedom are $q_i$,
  $i=0,\ldots, N-1$. In the limit as $k_c \to \infty$, we
  have both constraints, $q_{0}=0$ and $q_{N}=L$, in both
  models, making it obvious that they are identical.}

The system \eqref{eq:7v3} can be solved in an analogous way,
  by means of a perturbative expansion in the asymmetry
  $\xi$ and the pulling velocity $v_{p}$. The result is
\begin{eqnarray}
x_{i}=\ell &+& \frac{\xi N k_c \overline{\delta f}(\ell)-v_{p}\gamma k_c\dfrac{[3U''(\ell)+k_c(N-1)]N(N+1)}{6[Nk_c+U''(\ell)]}}{U''(\ell)[Nk_c+U''(\ell)]}  \nonumber
\\
&+& \frac{v_{p}\gamma k_c i\left(i-1+\dfrac{2U''(\ell)}{k_c}\right)-2\xi [Nk_c+U''(\ell)] \delta f_{i}(\ell)}{2U''(\ell)[Nk_c+U''(\ell)]}.
\label{eq:25}
\end{eqnarray}
Again, we can reobtain \eqref{eq:23} taking the infinite stiffness
limit in \eqref{eq:25}. Although the final solution for the extension
is different from the previous one, when we look for the critical
velocities and make the approximation $\ell_c \simeq \ell_b$ we get
the same analytical results for them.

Our main conclusion is that the existence of a set of critical
velocities, setting apart regions where the first unit to unfold is
different, is not an artificial effect of the limit $k_c \to \infty$.
Indeed, at the lowest order, all the versions of the studied model,
independently of the value of the stiffness and the location of the
spring, give the same critical velocities. This robustness is an
appealing feature of the theory, and makes it reasonable to seek this
phenomenology in real experiments.

\section{Shape of the bistable potentials}
\label{sec:potentials}

Different shapes for the double-well potentials have been
  considered in the literature, mainly simple Landau-like quartic
  potentials to understand the basic mechanisms underlying the
  observed behaviours \cite{PCyB13,BCyP15,GMTCyC14} and more complex potentials when trying to
  obtain a more detailed, closer to quantitative, description of the
  experiments \cite{BCyP15,BGUKyF10,BHPSGByF12,BCyP14}. For the sake of concreteness, we
restrict ourselves to the proposal made by Berkovich et
al.~\cite{BGUKyF10,BHPSGByF12}. Therein, the free energy of a
module is represented by the sum of a Morse
potential, which mimics the enthalpic minimum of the folded
state, and a worm-like-chain (WLC) term \cite{MyS95}, which accounts for the entropic
contribution to the elasticity of the unfolded
state. Specifically, the free energy is written as
 \begin{equation}
 U(x)=U_0 \left[ \left( 1 - e^{-2b\frac{x-R_c}{R_c}} \right)^2 -1 \right] + \frac{k_BT}{4P}L_c
 \left( \frac{1}{1-\frac{x}{L_c}} -1 -\frac{x}{L_c}+\frac{2x^2}{L_c^2} \right).
 \end{equation}

 This shape has shown to be useful for some pulling experiments with
 actual proteins as titin I27 or ubiquitin
 \cite{BGUKyF10,BHPSGByF12}. Therein, each parameter has a physical
 interpretation. First, in the WLC part, we have: (i) the contour
 length $L_c$, which is the maximum length for a totally extended
 protein, and (ii) the persistence length $P$, which measures the
 characteristic length over which the chain is flexible. Both of them,
 $L_c$ and $P$, can be measured in terms of the number of amino acids
 in the chain. Second, for the Morse contribution, we have: (iii)
 $R_c$, which gives the location of the enthalpic minimum and (iv)
 $U_0$ and $b$, which measure the depth and the width (in a
 non-trivial form) of the folded basin. The stability threshold
 $\ell_b$ cannot be provided as an explicit function of the parameters
 in Berkovich et al.'s potential. However, we  can always estimate it
 numerically, solving $U''(\ell_b)=0$ for a specific set of
 parameters.

As we anticipated in section \ref{sec:orig}, here we will focus in a
very specific configuration where only the first module is different
from the rest. Consistently, we use $U(x)$ to represent the free
energy of each of the identical modules, and $U_{1}(x)$ for that of
the first one. In particular, we consider that the first unit has a
slightly different contour length, $L_c + \Delta$. Therefore, we can
linearise $U_{1}(x)$ around $U(x)$, using the natural asymmetry
parameter $\xi = \Delta/ L_c\ll 1$. Therefore,
\begin{equation}
U_{1}'(x;L_{c}+\Delta) \simeq U'(x;L_{c}) + \xi \delta f_{1}(x;L_{c}),
\end{equation}
where 
\begin{equation}
\label{eq:deltaf}
\delta f_{1}(x;L_{c}) \equiv L_c \frac{\partial U'(x;L_{c})}{\partial L_{c}}=-\frac{k_B T}{2P} \left[ \frac{\frac{x}{L_c}}{\left( 1-\frac{x}{L_c} \right)^3} +\frac{2x}{L_c} \right].
\end{equation}
This linearisation is useful for the direct application of our theory
to  some  engineered systems, see the next section.

\section{Experimental prospect}
\label{sec:experiments}

In the experiments, the observation of the unfolding pathway is not
trivial at all.  The typical outcome of AFM experiments is a
force-extension curve (FEC) in which the identification of the
unfolding events is, in principle, not possible when the modules are
identical. Thus, in order to test our theory, molecular engineering
techniques that manipulate proteins adding some extra structures, such
as coiled-coil \cite{LSyM14} or Glycine \cite{SKSNyR03} probes, 
come in handy. For instance, a polyprotein in which all the modules
except one have the same contour length may be constructed in this
way. A reasonable model for this situation is a chain with modules
described by Berkovich et al.~potentials with the same parameters for
all the modules, with the exception of the contour length of one of
them. According to our theory, a critical velocity emerges
\eqref{eq:vc} and it may be observed because the unfolding of the unit
which is different can be easily identified in the FEC, see below.

Let us consider an example of a possible real experiment for a polyprotein with $N=10$ modules. We characterise the
modules by Berkovich et al.~potentials with parameters,
\begin{subequations}\label{eq:parameter}
\begin{align}
 &P=0.4\text{ nm}, & &L_c=30\text{ nm}, & &R_c=4\text{ nm}, \\
 &b=2, & &U_0=100\text{ pN}\,\text{ nm}, & &k_BT=4.2\text{ pN}\,\text{ nm},
\end{align}
\end{subequations}
and the friction coefficient $\gamma=0.0028$pN$\,$nm$^{-1}$s
\cite{BGUKyF10}. We call this system M$_{10}$: since all the modules
are equal in M$_{10}$, it is not a very interesting system from the
point of view of our theory.  Nevertheless, we can consider a mutant
species M$'_{10}$ that is identical to M$_{10}$ except for the module
located in the first position (the fixed end), which has an insertion
adding $\Delta$ to its contour length. Our theory gives an estimate
for the critical velocity $v_c$ by using \eqref{eq:vc}.

\begin{figure}
\centering
\includegraphics[width=0.89 \textwidth]{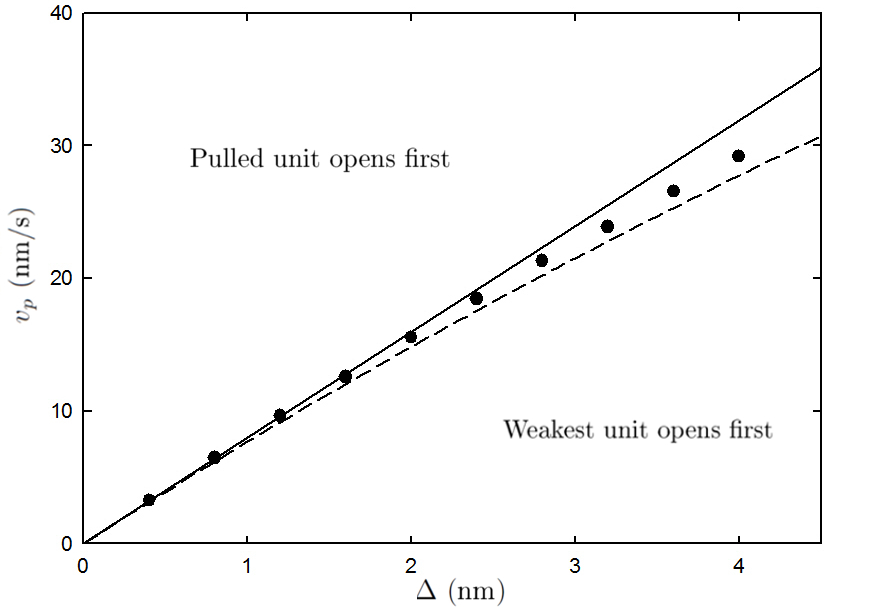}
\caption{\label{fig:phase} Critical velocities for M$'_{10}$
  systems. The parameter $\Delta$ stands for the additional contour
  length of the first module. Numerical values (circles) are compared
  with two theoretical estimates: ``complete'' (dashed line) and linear (solid line).}
\end{figure}

In Fig.~\ref{fig:phase}, we compare the theoretical estimate for the
critical velocity with the actual critical velocity obtained by
integration of the dynamical system \eqref{eq:7v3}. Specifically, we
have considered a system with spring constant $k_c=100$pN/nm. The
numerical strategy to determine $v_c$ has been the following: starting
from a completely folded state we let the system evolve obeying
\eqref{eq:7v3}, with a ``high'' value of $v_p$ (well above the
critical velocity), up to the first unfolding. We tune $v_p$ down
until it is observed that the first module that unfolds is the weakest
one: this marks the actual critical velocity. There are two
theoretical lines: the solid line stems from the rigorous application
of \eqref{eq:vc}, with $\delta f_{1}$ given by \eqref{eq:deltaf}, and
$v_{c}$ is a linear function of $\xi$, whereas the dashed line
corresponds to the substitution in \eqref{eq:vc} of
$\xi \delta f_{1}(x)$ by $U_{1}'(x;L_{c}+\Delta)-U'(x;L_{c})$, without
linearising in the asymmetry $\xi$.  Note the good agreement between
theory and numerics, specially in the ``complete'' theory where, for
the range of plotted values, the relative error never exceeds
5$\%$. Interestingly, the computed values of the critical velocity lie
in the range of typical AFM pulling speeds, from $10$ nm/s to $10^4$
nm/s \cite{AyA06}.

Below $v_{c}$, it is always the weakest unit that unfolds first. Above
$v_c$, the unit that unfolds first is the pulled one. For the sake of
concreteness, from now we consider an specific molecule M$'_{10}$
fixing $\Delta=2$ nm. Using the linear estimation \eqref{eq:deltaf} in
\eqref{eq:vc}, we get a critical velocity $v_c \simeq 16$ nm/s that,
as stated above, is in the range of the typical pulling speeds in AFM
experiments.

\begin{figure}
\centering
\includegraphics[width=0.89 \textwidth]{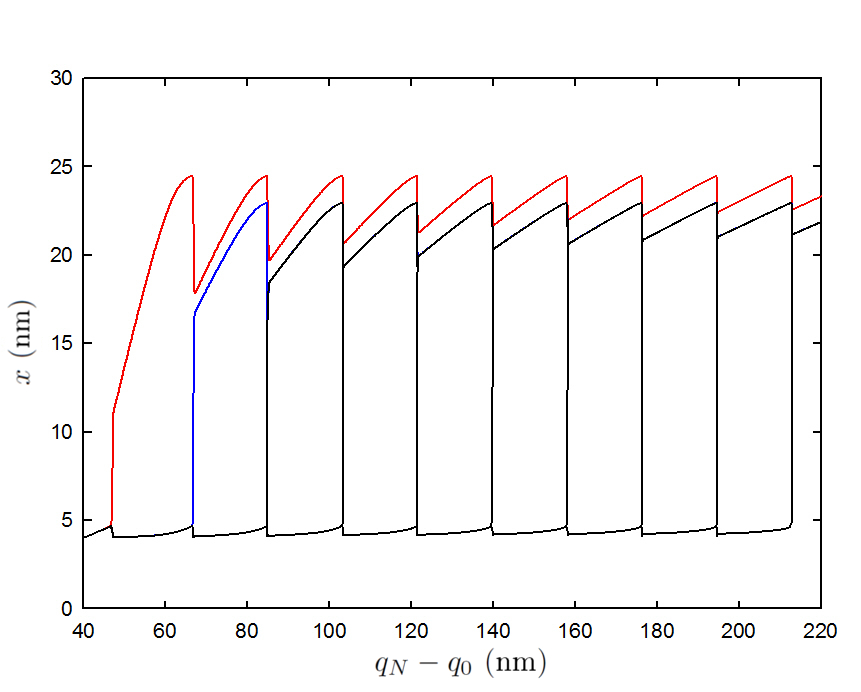}
\includegraphics[width=0.89 \textwidth]{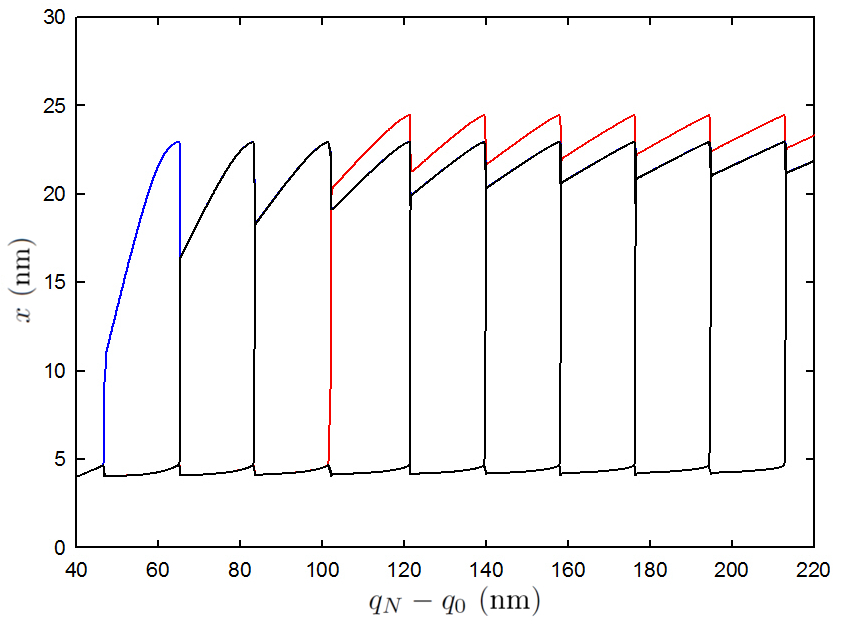}
\caption{\label{fig:3} Evolution of the extensions of the different
  units as a function of the length of the system $q_{N}-q_{0}$ in a
  pulling experiment. The potential parameters are given in
  Eq.~\eqref{eq:parameter}, and the pulling speeds are $v_{p}=10$ nm/s
  $<v_c$ (top) and $v_{p}=22$ nm/s $>v_c$ (bottom). The stiffness is
  $k_c= 100$ pN/nm, which lies in the range of typical AFM values.
  The red line corresponds to the weakest unit and the blue line to
  the pulled one.  }
\end{figure}

In Fig.~\ref{fig:3}, we plot the extension of each unit vs the total
extension $q_N-q_0$ in our notation ($X$ in
Fig.~\ref{fig:sketch-experiment}), which is a good reaction coordinate
\cite{ACDNKMNLyR10}. We have numerically integrated
Eqs.~\eqref{eq:7v3} for two values of $v_{p}$: one below and one above
$v_c$, namely $v_p=10$ nm/s and $v_p=22$ nm/s. The red trace stands
for the weakest unit extension whereas the blue one corresponds to the
pulled module. We can see that, for $v_p=10\text{ nm/s}<v_c$, the
first unit that unfolds is the weakest one, whereas for
$v_p=22\text{ nm/s}>v_c$ that is no longer the case. Specifically, the
first unit that unfolds is the pulled one, and the weakest unfolds in
the 4-th place.

\begin{figure}
\centering
\includegraphics[width=0.89 \textwidth]{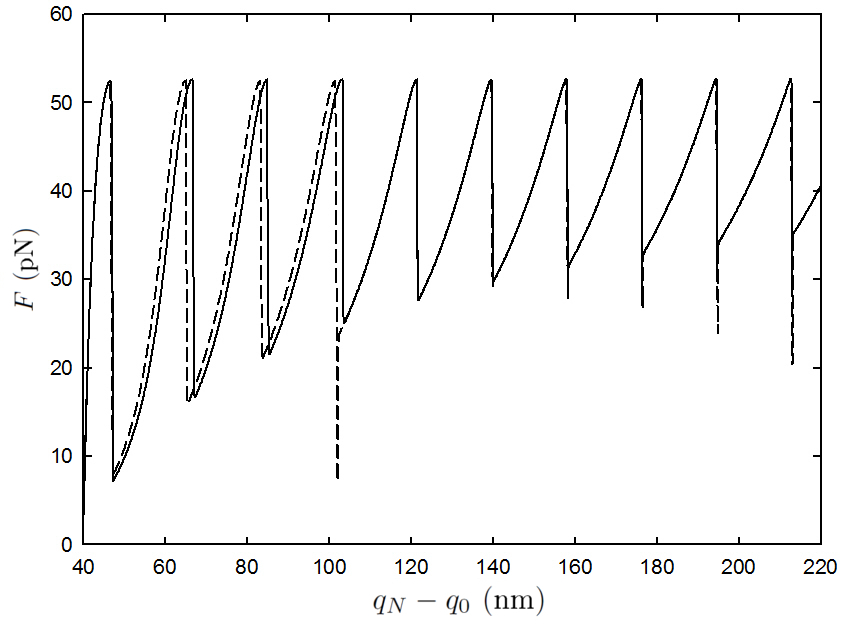}
\includegraphics[width=0.89 \textwidth]{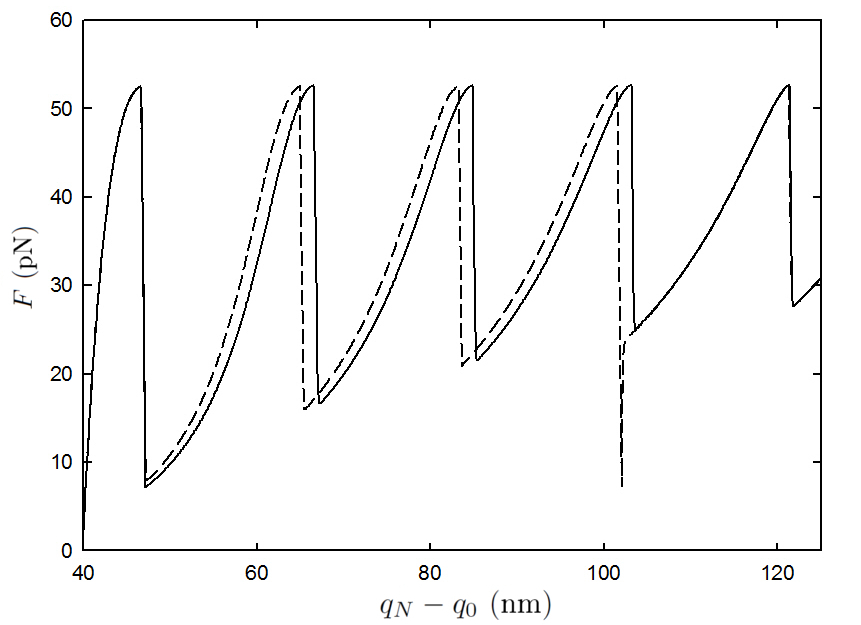}
\caption{\label{fig:4} Top: FEC for the pulling experiment in
  Fig.~\ref{fig:3}. Two pulling speeds are considered, specifically
  $v_{p}=10$ nm/s (subcritical, solid) and $v_{p}=22$ nm/s
  (supercritical, dashed). Bottom: zoom of the region of interest,
  showing the shift between the peaks stemming from the increased
  contour length of the mutant module.  }
\end{figure}

The plots in Fig.~\ref{fig:3} are the most useful in order to detect
the unfolding pathway of the polyprotein. Unfortunately, they
are not accessible in the real experiments, for which the
typical output is the FEC. Thus, we have
also plotted the FEC in order to bring to
light the expected outcome of a real experiment. In Fig.~\ref{fig:4}, we show the FEC for the two
considered velocities in the same graph (solid line for the lower speed
and dashed line for the higher one).

The FECS in Fig.~\ref{fig:4} are superimposed until the first force
rip, which corresponds to the first unfolding event (that of the
mutant module for the slower velocity and that of the pulled unit for
the faster one). As the mutant unit has a longer contour length than
the rest, a shift between the curves in the next three pikes is found,
because the effective contour length of the polyprotein has an extra
contribution of $2$ nm. Reasonably, for the higher velocity, this
shift disappears when the mutant module unfolds, and the curves are
once again superimposed. This plot clearly shows how the existence of
a critical velocity  in a real experiment could be sought.

\section{Conclusions}
\label{sec:conclusion}

We have provided a useful theoretical framework in the context of
modular proteins or, in general, of biomolecules comprising several
units that unfold (almost) independently. Therein, according to our
theory, it should be possible to find the emergence of a set of
critical velocities which separate regions where the first module that
unfolds is different. Although we focus on the biophysical application
of the theory, it is worth highlighting that similar models are used
in other fields. Many physical systems are also ``modular'', since
they comprise several units \cite{BZDyG16}, and thus a similar
phenomenology may emerge.  Some examples can be found in studies of
plasticity \cite{MyV77,PyT02}, lithium-ion batteries
\cite{DJGHMyG10,DGyH11} or ferromagnetic alloys \cite{BFSyG13}.

The development of our theory has shown that the position and value of
the elastic constant $k_c$ of the length-controlled device is roughly
irrelevant for the existence and value of the critical velocities. The
derived expressions for the critical velocities are, to the lowest
order, independent of the spring position and
stiffness. Notwithstanding, our theory completely neglects the noise
contributions and thus the units unfold when they reach their limit of
stability, that is, at the force for which the folded basin
disappears. This is expected to be relevant for biomolecules that
follow the \textit{maximum hysteresis path}, using the same
terminology as in \cite{BZDyG16,BCyP15}, completely sweeping the
metastable part of the intermediate branches of the FEC.

The numerical integration of the evolution equations for a realistic
potential point out that our proposal for an experiment is, in
principle, completely feasible. Therefore, our work encourages and
motivates new experiments, in which the predicted features about the
unfolding pathway of modular biomolecules could be observed. Finally,
the discussion in the previous paragraph on the relevance of thermal
noise makes it clear that an adequate choice of the biomolecule is a
key point when trying to test our theory.

\begin{acknowledgement}
  We acknowledge the support of the Spanish Ministerio de
  Econom\'{\i}a y Competitividad through Grant FIS2014-53808-P. Carlos
  A. Plata also acknowledges the support from the FPU Fellowship
  Programme of the Spanish Ministerio de Educaci\'on, Cultura y
  Deporte through Grant FPU14/00241. We also thank
  Prof. P.~Marszalek's group for sharing with us all their knowledge
  and giving the opportunity to start exploring AFM experiments during
  a research stay of Carlos A.~Plata at Duke University in summer
  2016, funded by the Spanish FPU programme. Finally, we thank the
  book editors for their organisation of the meeting at the excellent
  BIRS facilities in Banff, Canada, in which a preliminary version of
  this work was presented in August 2016.
\end{acknowledgement}

%
%
%

\end{document}